\documentclass[12pt,preprint]{aastex}

\usepackage{graphicx}
\usepackage{epstopdf}
\usepackage{stmaryrd}
\usepackage{nicefrac}
\usepackage{color}

\def\apj{{ApJ}}%
\def\aap{{A\&A}}%
\def\jcap {{J. Cosmology Astropart. Phys.}}%
\def\mnras{{MNRAS}}%
\def\prl{{Phys.~Rev.~Lett.}}%
\def\simle{\lower 2pt \hbox {$\buildrel < \over {\scriptstyle \sim }$}}
\def\simge{\lower 2pt \hbox {$\buildrel > \over {\scriptstyle \sim }$}}

\begin{document}

\title{Ultra-High Energy Cosmic Rays from Centaurus A: \\ Jet Interaction with Gaseous Shells}

\author{Gopal-Krishna\altaffilmark{1},   
Peter L.\ Biermann\altaffilmark{2,3},
Vitor de Souza\altaffilmark{4},
Paul J.\ Wiita\altaffilmark{5,6}}
\altaffiltext{1}{National Centre for Radio Astrophysics, Tata Institute of Fundamental Research, Pune, 411007, India; e-mail: krishna@ncra.tifr.res.in}
\altaffiltext{2}{Max-Planck-Institute for Radioastronomy, Auf dem H{\"u}gel 69, 53121 Bonn, Germany}
\altaffiltext{3}{Also at: Dept.\ Physics \& Astronomy, Univ.\ Bonn, Germany; Dept.\ Physics \& Astronomy, Univ.\ Alabama, Tuscaloosa, AL; Dept.\ Physics, Univ.\ Alabama at Huntsville, AL; Dept.\ Physics., Karlsruher Institut f{\"u}r Technologie, Karlsruhe, Germany}
\altaffiltext{4}{Instituto de F\'{\i}sica de S\~{a}o Carlos, Universidade de S\~{a}o Paulo, Av.\ Trabalhador  S\~{a}o-carlense 400, Centro, CEP 13566-590, S\~{a}o Carlos, Brazil}
\altaffiltext{5}{Department of Physics, The College of New Jersey, PO Box 7718, Ewing, NJ 08628}
\altaffiltext{6}{Department of Physics and Astronomy, Georgia State University, PO Box 4106, Atlanta, GA 30302--4106}


%
         
\begin{abstract}
Ultra high energy cosmic rays (UHECRs), with energies above $\sim 6
\times 10^{19}$eV, seem to show a weak correlation with the distribution of matter relatively near to us  in the universe. It has earlier been proposed that UHECRs could be accelerated in either  the nucleus or the outer lobes of the nearby radio galaxy Cen A.
We show that UHECR production at a spatially intermediate location about 15 kpc northeast from the nucleus, where the jet emerging from the nucleus is observed to strike a large star-forming  shell of gas, is a plausible alternative. A  relativistic jet is capable of accelerating lower-energy heavy seed CRs to 
UHECRs on timescales comparable to the time it takes the jet to pierce  the large gaseous cloud.
In this model many cosmic rays arising from a starburst, with a composition enhanced in heavy elements near the knee region around PeV,   are boosted to ultra-high energies by the relativistic shock of a newly oriented jet.  This model matches the overall spectrum shown by the Auger data and also makes a prediction for the chemical composition as a function of particle energy. We thus predict an observable anisotropy in the composition at high energy in the sense that lighter nuclei should preferentially be seen  toward the general direction of Cen A.  Taking into consideration the magnetic field models for the Galactic disk and a Galactic magnetic wind, this scenario may resolve the discrepancy between HiRes and Auger results concerning the chemical composition of UHECRs.
\end{abstract} 


\section{Introduction}

To understand the origin of cosmic rays (CRs), it is important 
to distinguish
between the lower energy CRs which can be contained within
the magnetic field of our Galaxy and thus have energies of
up to about $3 \times 10^{18}$ eV for heavy
nuclei, and those that are even more energetic.
The bulk of the CRs below that energy can be explained by 
supernova explosions,
while the extremely energetic ones probably originate from either 
some class of Active Galactic Nuclei \citep{ginzburgsyrovatskii64,biermannstrittmatter87}
or some extreme type of stellar activity such as gamma-ray bursts~\citep{waxman95}.
Indeed, the spectrum of CRs
shows a kink near $3 \times 10^{18}$ eV, matching the expectation that 
their origin changes around this energy threshold.

Stellar explosions can account for the flux, spectrum,
particle energy and chemical composition of the less energetic CRs,
considering that all very massive
stars explode into their pre-existing winds~\citep[e.g.][]
{prantzos84,stanevetal93,meyer97}.  
Further quantitative confirmation of this picture has now emerged from detailed observations of
cosmic ray electrons and positrons, as well as the
WMAP haze ~\citep{biermannetal09b,biermannetal10}. The supernova origin of Galactic 
cosmic rays may
lead us to an understanding of the seed particle population ~\citep{biermannetal09a} on which active galactic nuclei energizing radio galaxies can operate their acceleration processes.

The origin of ultra high energy cosmic rays (UHECRs) is still an unresolved
issue, but a few clues have begun to emerge. Although their arrival directions are nearly
isotropic, a general correlation with the distribution of matter 
has been noted by the Auger observatory \citep{stanevetal95,auger08a,auger08b}, although it is disputed by the HiRes observatory~\citep{hires08,hires09}.
In particular, there may be excess events with arrival directions close to the
nearby radio galaxy Centaurus A~\citep{auger08a,auger09}. There are contradicting claims 
from  experiments as to
whether the UHECR events are heavy nuclei \citep[Auger --][]{abraham09} or purely protons \citep[HiRes --][]{abbasietal09}. 
Both possibilities need to be explored.

In a picture where UHECR energies are attained by a single kick up from a seed population~\citep{gallantachterberg99} through the
action of a relativistic jet, these events can indeed involve heavy nuclei~\citep{biermannetal09a}.
 In such a scheme the seed particles are the CRs 
near the spectral knee~\citep{stanevetal93}
and the relativistic shock is very likely to 
 arise from a jet carving out a new channel after being launched from a primary central black hole
that has been reoriented following the merger of the nuclear black holes of two merging galaxies
~\citep{gergelybiermann09}. In this scenario all the UHECR
particles are a mix of heavy nuclei, and the spectrum in ~\citet{stanevetal93}
 actually gives an adequate fit to the Auger data~\citep{biermannetal09a}.  See Fig.\ 1. The
sky distribution is easily isotropized by the intergalactic magnetic 
fields~\citep{dasetal08};
for the case of heavy nuclei one is even confronted with the possibility of excessive scattering~\citep{biermannetal09a}. This picture 
also allows the incorporation of 
the Poynting flux limit  \citep{lovelace76}: the 
particles to be
accelerated must remain confined within the jet diameter. This condition
translates into a
lower limit for the jet power, allowing most UHECR particles to 
originate from the jet interacting with lower energy CRs produced in the starburst in the central 
region of Cen A.

Therefore we explore a scenario based on the observed head-on encounter of the Cen A jet with magnetized interstellar clouds~\citep{gksaripalli84,kraftetal09,gkwiita10}
from which UHECR acceleration 
ensues. A distinctly appealing aspect of this 
proposal is that the postulated jet-cloud interaction is actually 
observed within the northern lobe of Cen A, whereby the jet is seen to be
disturbed, bent westward and possibly disrupted temporarily~\citep{morgantietal99,oosterloomorganti05}.
 Since any supersonic flow reacts to a disturbance with
shock formation, this in turn could cause particle acceleration. Note 
that the Fermi/LAT error circle for the peak of the gamma-ray emission~\citep{fermiLAT10} 
encompasses the jet-cloud interaction region, at the base of the  northern middle lobe of Cen A, about 15 kpc from the nucleus.  

\section{Acceleration in Cen A from a jet interacting with gaseous shells}

The key point is that the interaction of the northern jet  with a gaseous 
shell in the northern middle lobe has clearly been seen~\citep{oosterloomorganti05,kraftetal09}  
and massive star formation is revealed at the location of the interaction by the GALEX UV image~\citep{kraftetal09}.
Although other mechanisms can bend and disrupt radio jets, only a jet-shell interaction can explain the variety of data (radio, HI, UV, 
X-rays) for Cen A~\citep{kraftetal09}. It has also been argued that the oft-debated peculiar 
morphology of the northern middle radio lobe can be readily understood 
in terms of the same jet-shell collision~\citep{gkwiita10}. 

An important aspect of the basic acceleration physics to be stressed is that 
when particles are accelerated in a shock propagating parallel to the 
magnetic field lines, the maximum particle energy $E_{max}$ is given by~\citep{hillas84,ginzburgsyrovatskii64,stanev04}
$E_{max} \; = \; e \, Z \, \beta_{sh} \, R_{B} \, B$,
where $e$ is the elementary electric charge, $Z$ is the
numerical charge of the particle, $\beta_{sh}$ is the shock speed in units
of the speed of light, the available length scale is $R_{B}$,
and the strength of the magnetic field is $B$.
However, when the shock propagation is highly oblique, the corresponding limit~\citep{jokipii87,meli06} becomes
\begin{equation}
E_{max} \; = \; e \, Z \, R_{B} \, B,
\end{equation}
which is independent of the shock velocity. Invoking
relativistic shocks obviously adds an additional factor of $\gamma_{sh}$,
the shock's Lorentz factor~\citep{gallantachterberg99}.  Losses will
curtail this maximum attainable energy~\citep{hillas84,biermannstrittmatter87}. 

We now focus on the particle acceleration due to the observed interaction of
the jet with shells of fairly dense gas. Cen A has long been known to 
have a number of stellar shells, located in the vicinity of both the 
Northern and Southern lobes~\citep{malinetal83}. Some of these shells have later been found 
to contain large amounts of
dense atomic~\citep{schiminovichetal94}  and even molecular~\citep{charmandarisetal00}
 gas ($\sim 7.5 \times  10^8~$M$_{\odot}$).
These shells are generally thought to have originated from the
merger of a massive elliptical with a disk galaxy~\citep{quinn84}, 
very probably the same merger that gave rise to
the peculiar overall appearance of this large elliptical galaxy marked 
by a striking dust lane. Radio maps reveal that the northern
jet has encountered such shells at distances of 3.5 and 15 kpc from the 
core, and flared up each time to the same side, thereby forming the 
northern-inner and the northern-middle lobes~\citep{gksaripalli84,gkwiita10}.
Simulations of such collisions indicate the 
formation of strong shocks where the jets impinge
upon gas clouds~\citep[e.g.][]{choietal07}. 

We must ask whether the maximum observed particle energies, of order
$10^{21}$ eV, are actually attainable in such interactions.
Accelerating particles to such copious energies requires that the Larmor 
motion of a particle must fit within the
gaseous cloud, both before and after the shock that forms inside the
cloud by interaction with the impinging relativistic jet. This leads to 
the condition
$E_{max} \; \simle \; e \, Z \, B_{cl} \, R_{cl}$,
also called the Hillas limit~\citep{hillas84}, which is a general requirement to produce
UHECR via shocks.

Adopting the very reasonable parameter values of 3 kpc for $R_{cl}$, the approximate
observed size of the HI shell found in the Northern Middle Lobe of Cen A~\citep{oosterloomorganti05,gkwiita10},
 and $3 \times 10^{-6}$ Gauss for the magnetic field, it follows that the energy 
must remain below $Z \times 10^{19}$eV.
Since particles are observed up to about $3 \times 10^{20}$
eV~\citep{birdetal94}, this implies that heavy nuclei, such as Fe, are 
much preferred for this mechanism
to suffice; however, if a stronger magnetic field were present, 
this would ease the requirement on the abundances and allow for CRs to be accelerated
to even higher energies. The magnetic field in the shell is not well constrained, but the 
required value is  modest.

The Hillas limit condition~\citep{hillas84} mentioned above can be expressed another way~\citep{lovelace76}.
Taking the energy needed for particle acceleration to derive from a jet,
we can connect the time-averaged energy flow along the
jet with the condition that the accelerated particles are contained
within the jet diameter,
\begin{equation}
L_{jet} \; \simge \; 10^{47} \, {\rm erg~s}^{-1} \, f_{int} \,{\left(
\frac{E_{max}}{Z \times 10^{21} {\rm eV}} \right)}^{2},
\end{equation}
where $f_{int}$ is an intermittency factor describing the temporal
fluctuations of the energy outflow. Equality in this critical expression
would imply that the energy flow in the jet is an entirely electromagnetic Poynting flux,
an unrealistic
extreme scenario. For Cen A we require both an intermittency factor $< 1$,
and presumably also heavy nuclei, e.g., $Z \simeq 26$. 
We find $f_{int} \, \simle \, 0.75$ in
order to match the kinetic jet power, which has been argued to be $L_{jet} \simeq 10^{43}$
erg s$^{-1}$ through several different approaches \citep[][and references therein]{whysongantonucci03,fermiLAT10,kraftetal09}.   The recent HESS observations
of Cen A \citep{hess09} detected an ultra-high energy ($> 250$ GeV) photon luminosity of
only $\simeq 2.6 \times 10^{39}$ erg s$^{-1}$, but the entire photon luminosity in gamma-rays ($>100$ keV)
is $\sim 2 \times 10^{42}$ erg s$^{-1}$, and thus also consistent with $L_{jet} \simeq 10^{43}$ erg s$^{-1}$.

We next examine whether the inferred luminosity of UHECRs is indeed
attainable.  Assuming the observed
spectrum of the jet corresponds to a CR particle spectrum of about
$E^{-2.2}$, this leads to the requirement that the observed power in UHECR
particles must be multiplied by a factor of about 200 in order to integrate 
over the power-law
spectrum. The data then require a luminosity of about $10^{42}$ erg 
s$^{-1}$, still below the inferred
jet power
of $10^{43}$ erg s$^{-1}$ for Cen A~\citep{whysongantonucci03, fermiLAT10}. Thus, 
we could allow
for a duty cycle of 0.1, and still have adequate jet power. So 
the jet's interactions with a dense
cloud are capable of powering the observed UHECRs.
Another way of asking the same question is, can a jet actually catch a sufficient number of particles from the knee region with energies near PeV and accelerate them to the ankle region near EeV to ZeV? 
Assuming that the energy density of CRs in the starburst region is about 100 times what we have in our galaxy, the particle density near and above $10^{15}$ eV is about $10^{-17}$ per cc.  
If through the non-steadiness of the jet these CRs are caught at the same rate by a kpc scale jet
having an opening angle of, say, $5^{\circ}$, the cross-section of $\sim 10^{41.5}$
cm$^2$ implies a rate of $10^{35}$ particles accelerated per second.
Pushing them to UHECR energies gives an energy turnover of order $10^{42}$ erg s$^{-1}$ just for the energies above $10^{18.5}$ eV,  again quite sufficient.

Third, we need to check whether enough time is available for the
particles to be
accelerated. A jet encounter with such a large cloud would last for at least 
$10^{4}$ yrs~\citep{choietal07}.
A shock in either the external or the internal medium would take some small multiple of the Larmor time scale at the
maximum energy of a few times $10^{4.3}$ yrs, to complete the
acceleration process. The two relevant time scales, for
transit and acceleration, seem consistent 
within the scope of our broad estimates. 

Lastly, we need to check whether the time scales are long enough so
that the time window for possible detection of the UHECR source is not
too brief.  The time scales for particle
acceleration and the jet-cloud encounter are somewhere between $10^{4}$ 
and $10^{5}$ yr. The times for the jet to transit a shell and then to move on to the next shell appear to be in a ratio of about 1 to 10. Therefore, a duty cycle, $f_{int}$,  of about 0.1, which is easily allowed for by the above calculation, is actually necessary to maintain a quasi-continuous output of accelerated particles.

\section{Consequences of the jet/cloud acceleration scenario}

Having shown that the basic model is viable, we now consider some of its consequences.

First we note that the jet may still be mildly precessing after the 
episode of
the merger of black holes~\citep{gergelybiermann09}, in the 
aftermath of the merger of the
elliptical and spiral galaxies comprising Cen A~\citep{fisrael98}.
Also, the gaseous shell may have its own motion, also due to the 
preceding
merger of the two galaxies. This would naturally explain the observed
multiple bendings and flarings of the northern jet in Cen A~\citep{gksaripalli84,gketal03,gkwiita10}. 
Both effects would expose continuously fresh material to the action of the jet,
but are not an essential requirement for our model.


Second, the transport and scattering of the particles along the way might
smooth out any variability even if Cen A were the only significant
source of UHECRs in our part of the universe.  Such variability might explain the
inconsistencies between Auger and HiRes results~\citep{abraham09,hires09}.
The magnetic field at the site of origin is locally enhanced
by the Lorentz factor of the shock, possibly between 10 and 50 \citep[e.g.,][]{biermannetal09a}.  That could imply the shortest possible variability time $\tau_{var} \simeq 100 \, \tau_{var, 2}$ yrs,
taking a high Lorentz factor of 50. 
The scattering near the Earth needed
to attain near isotropy in arrival directions
requires a relatively strong magnetic field within the distance equal to  
$c \tau_{var}$. So the containment of Fe particles of up to $3 \times 10^{20}$ eV
would imply an energy content near Earth of 
$E_{B, var} \; \simge \; 6 \times 10^{51} \, \tau_{var, 2}^{+1} \, {\rm erg}$.
Interestingly, this total energy approaches the energy of a hypernova ($10^{52}$ erg).
However, there is currently no evidence for such a region surrounding the Sun.




Finally, we have to follow through with the deduction from the Poynting flux limit \citep{lovelace76}, that the highest energy events can only be heavy nuclei if they come from Cen A.  This limit requires that all particles caught by a shock in the jet have $E/Z$ less than or equal to that of Fe at $10^{20.5}$ eV, the highest energy event  yet seen; let us assume initially, that this one event at $10^{20.5}$ eV is a factor of 3 below the real limit imposed by the acceleration site, the shock in the jet interaction region.  It follows that He above $10^{19.9}$, and CNO above $10^{20.1}$ eV are ruled out, but near $10^{19.7}$ eV both are possible.  We use the prescriptions of \citet{allard08} to define a photo-disintegration distance $\Lambda_{dis}$ for any nucleus and energy.  Averaging over some wiggles in the curves that cover both FIR and microwave backgrounds from the very early universe, we find that over the relevant energy and nucleus charge range an adequate approximation  is $\Lambda_{dis} \, = \, 10^{1.6} \, {\rm Mpc} \, (Z/Z_{Fe})/(E/10^{19.7} \, {\rm eV})^{2.6}$, which we use here to guide us.  There are two extreme scattering limits. In one limit, the isotropization of the events from Cen A is done in the intervening intergalactic medium (IGM). Cosmological MHD simulations by \citet{Ryu08} imply a Kolmogorov approximation; then $\Lambda_{trav} \, = \, 10^{1.6} \, {\rm Mpc} \, [(Z/Z_{Fe})/(E/10^{19.7} \, {\rm eV})]^{1/3}$.  However, this already leads to extreme losses of the heavy nuclei between Cen A and us.  So we 
consider the other limit, in which the UHECRs travel essentially straight from Cen A to us, and are isotropized in the magnetic wind of our Galaxy \citep{Everett08}.   Modeled values of the wind's magnetic field strength ($\sim 8~\mu$G) and radial scale ($\sim 3$ kpc) allow Fe, as well as all elements down to about Oxygen, to be scattered into isotropy; however, there is less effect on lower $Z$ elements. No other approach gave a reasonable fit to the data.  The losses due to the path traversed during the scattering are small.  A fit with this approach is shown in Fig.\ 1.  One could use other magnetic wind model numbers, closer to a Parker-type wind~\citep{Parker58}, but the essential results do not change.     Now we must ask, how can this be compatible with IGM models \citep{Ryu08,dasetal08,choryu09}?  Given the overall magnetic energy content in the IGM, scattering can be reduced if much of the overall magnetic energy is pushed into thin sheets \citep{biermannetal09c} and such substructure  plausibly arises from radio galaxies and galactic winds.  A second question is whether the magnetic field could also produce a systematic shift on the sky for UHECRs, in addition to scattering and isotropizing them.  Indeed, any Galactic magnetic field~\citep{beck96}, in the disk or in the foot region of a Galactic wind~\citep{stanev97,Everett08} would also produce a systematic shift relative to the central position of Cen A on the sky.  Since Cen A is not far from the sensitivity edge of the Auger array in the sky, it is quite possible that there is a shift for all events, especially at slightly lower energies.  The models of ~\citet{Zirakashvili96} show that angular momentum conservation and transport quickly generate a magnetic field component parallel to the galactic disk, which would shift particle orbits in a direction perpendicular to the disk, and possibly away from the center of symmetry.  

A testable prediction of this scenario then is that  a solid angle on  the sky 
containing half the UHECR events  towards Cen A should show the signature of lighter nuclei, hence larger fluctuations, compared to the events seen from the remaining part of the sky.  Since the main scattering also has a systematic component, the  center of this anisotropy may be shifted with respect to Cen A, so that the part of the sky with the largest fluctuations in the shower properties may be 
offset by up to a few tens of degrees from Cen A for $Z > 1$.  This effect might be strong enough so that in  some parts of the sky lighter elements might predominate over heavies and thus reconcile results from the Auger and HiRes experiments~\citep{abbasietal09,abraham09}.  This could  soon be checked with the growing data on UHECRs. If such a test were positive, it would unequivocally and simultaneously show that Cen A is the best source candidate, that scattering depends on the energy/charge ratio, and that the most energetic events are heavy nuclei.

PLB acknowledges discussions with J.\ Becker, L.\ Caramete, S.\ Das, T.\ Gaisser, L.\ Gergely, 
H.\ Falcke, S.\  Jiraskova, H.\ Kang, K.-H.\ Kampert, R.\ Lovelace, A.\ n, M.\ Romanova, D.\ Ryu, T.\ Stanev, and his Auger Collaborators, especially H.\ Glass. PLB also acknowledges the 
award of a Sarojini Damodaran Fellowship by the Tata Institute of 
Fundamental Research.  VdS is supported by FAPESP (2008/04259-0) and CNPq.\\

\vspace*{12pt}
\begin{figure}[h!]
\centering
\includegraphics[bb=0cm 0cm 20cm 15cm,viewport=0.0cm 0cm 19cm 
14cm,clip,scale=0.65]{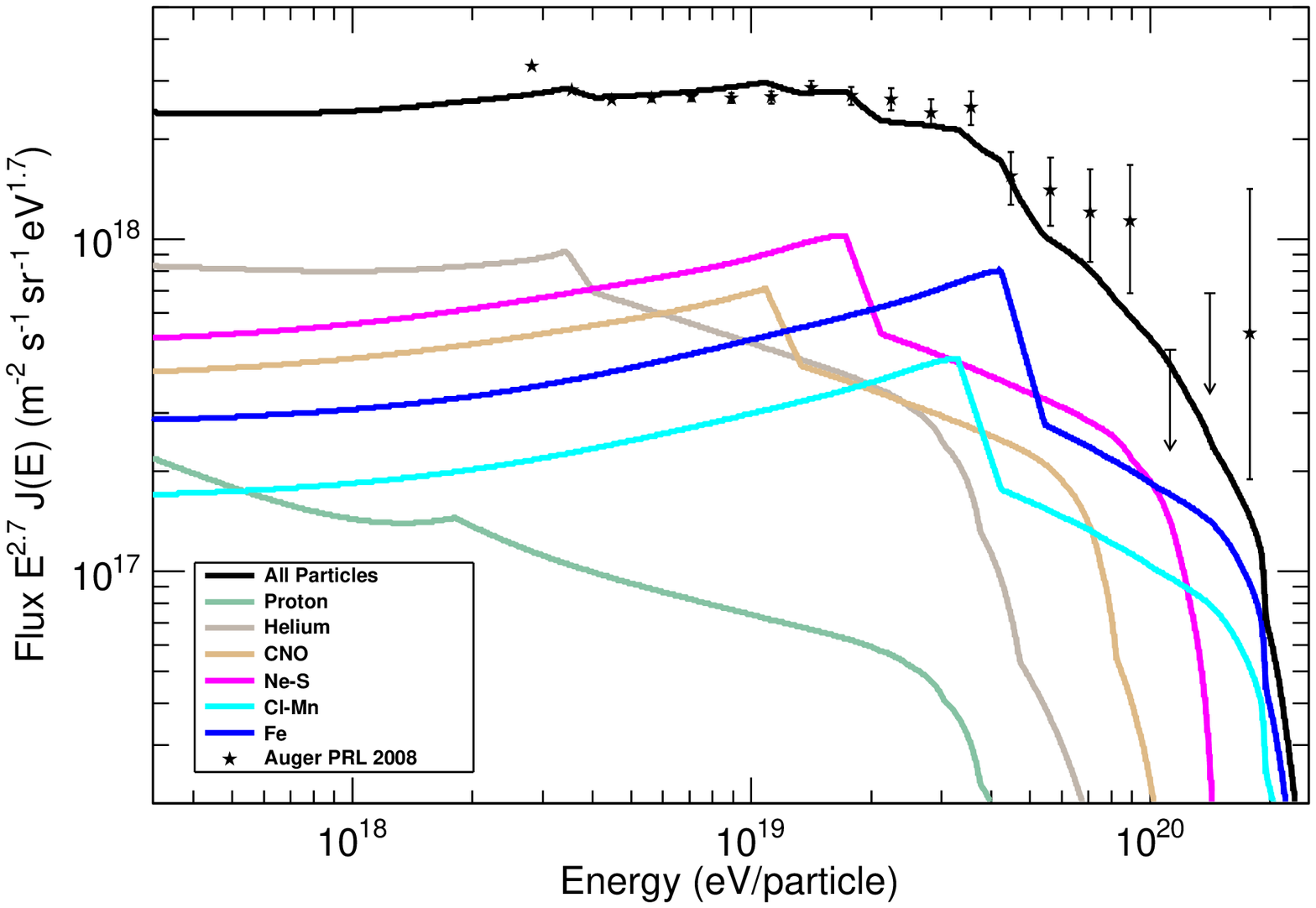}
\vspace*{0pt}
\caption{Testing the shift of the spectrum in paper \citep{stanevetal93} with Auger data,  using
the propagation calculations of \citep{allard08} with a 3.8 Mpc distance to Cen A and assuming all isotropizing is in the magnetic wind of our Galaxy \citep{Everett08}. Note that the break here is due to the MHD structure of massive star winds,  pushed to EeV energies by a highly relativistic shock.
A color version appears in the on-line journal.}
\end{figure}
\label{Spectrum} 

\end{document}